\documentclass[twocolumn,fleqn,10pt,showpacs]{revtex4}

\usepackage{amsmath}
\usepackage{amssymb}
\usepackage{graphicx,psfrag,epsfig}
\usepackage{afterpage}
\usepackage{mathrsfs,simplewick}

\newcommand{\be}{\begin{equation}}
\newcommand{\ee}{\end{equation}}
\newcommand{\bea}{\begin{eqnarray}}
\newcommand{\eea}{\end{eqnarray}}
\newcommand{\bean}{\begin{eqnarray*}} 
\newcommand{\eean}{\end{eqnarray*}}

\newcommand{\bm}[1]{\mbox{\boldmath $#1$}}

\newcommand{\ket}[1]{\vert #1 \rangle}

\def\lag{{\mathscr L}}
\def\!{\hat}
\def\slash{\rlap{/}}

\def\slash#1{\setbox0=\hbox{$#1$}               
        \dimen0=\wd0                            
        \setbox1=\hbox{/} \dimen1=\wd1          
        \ifdim\dimen0>\dimen1                   
        \rlap{\hbox to \dimen0{\hfil/\hfil}}    
        #1                                      
        \else                                   
        \rlap{\hbox to \dimen1{\hfil$#1$\hfil}} 
        /                                       
        \fi}                                    %

\begin{document}

\title{The Roots of the Standard Model of Particle Physics}

\author{P.J.~Mulders}
\email{mulders@few.vu.nl}
\affiliation{
Nikhef Theory Group and Department of Physics and Astronomy, 
VU University Amsterdam,\\
De Boelelaan 1081, NL-1081 HV Amsterdam, the Netherlands}

\begin{abstract}
We conjecture how the particle content of the standard model can emerge starting with a supersymmetric Wess-Zumino model in 1+1 dimensions ($d = 2$) with three real boson and fermion fields. Considering $SU(3)$ transformations, the lagrangian and its ground state are $SO(3)$ invariant. The $SO(3)$ symmetry extends the basic $IO(1,1)$ Poincar\'e symmetry to $IO(1,3)$ for the asymptotic fields requiring physical states to be singlets under the $A_4$ symmetry that governs the  $SO(3)$ embedding. This is linked to the three-family structure. For the internal symmetries of the asymptotic fields an $SU(2) \times U(1)$ symmetry remains, broken down as in the standard model. 
The boson excitations in $d = 4$ are identified with electroweak gauge bosons and the Higgs boson. Fermion excitations come in three families of leptons living in $E(1,3)$ Minkowski space or three families of quarks living in $E(1,1)$. 
Many features of the standard model now emerge in a natural way. The supersymmetric starting point solves the naturalness problem. The underlying left-right symmetry leads to custodial symmetry in the electroweak sector. In the spectrum one has Dirac-type charged leptons and Majorana-type neutrinos. The electroweak behavior of the naturally confined quarks, leads to fractional electric charges and the doublet and singlet structure of left- and right-handed quarks, respectively. 
Most prominent feature is the link between the number of colors, families and space directions. 

\end{abstract}

\maketitle

\section{Introduction}

The standard model of particle physics is highly successful incorporating besides gravity all known particles and their interactions. The theoretical framework is a renormalizable gauge theory even if the gauge symmetry is a rather ad hoc combination of an unbroken $SU(3)$ symmetry group with a spontaneously broken $SU(2)\otimes U(1)$ symmetry group for strong and electroweak interactions. It requires besides different coupling constants also a multitude of parameters governing the coupling with the Higgs sector, responsible for the electroweak symmetry breaking and the mixings and masses of quarks and leptons. Striking connections between the electroweak and strong sectors are the vanishing of baryon minus lepton number (the proton excess balances the electron excess), or within the electroweak sector the relation between $SU(2)$ and $U(1)$ coupling constants involving a weak mixing angle $\sin\theta_W \approx 1/2$ suggesting an $SU(3)$ embedding. There are strong indications that we must look beyond the standard model, such as the presence of dark matter in the universe and occasional indications of cracks in the model from precision experiments but for one obvious route, namely compositeness of quarks and leptons, there are so far no indications. Finally we mention the universality of the electroweak and strong forces for the three families of quarks and leptons as a striking feature. 

In this paper~\cite{conference}, I want to sketch a scenario that could provide a new starting point for looking at the roots of the standard model, even if there remain several loose ends that need to be looked at in detail and even if it might not affect existing results. We argue that a natural emergence of abovementioned striking features can be linked to the fact that in the asymptotic world the interactions are the electroweak ones (color is hidden) and there is a Poincar\'e symmetry $IO(1,3)$ with three space directions.

\section{The starting point}

As starting point we take {\em one} space dimension (1D world) with a $d=2$ Poincar\'e symmetry $IO(1,1)$. We take a field theoretical route rather than a string theoretical one. The Poincar\'e symmetry is central in the Hilbert space, with Hamiltonian and momentum operator generating time and space translations and boosts transforming among momentum eigenstates. With only one time and one space direction, states live in an $E(1,1)$ Minkowski space with coordinates $x^\mu$ and metric $x^2$. When appropriate, we will use light-cone components employing light-like vectors $n = n_+ = (1,1)$ and $\bar n = n_- = (1,-1)$, thus $a{\cdot}\bar n = a^+$ and $a{\cdot}n = a^-$. 
The quantum states $\ket{k}$ in the free theory are associated with the modes of field oscillations around the classical (minimum energy) solution, for free fields eigenstates of the momentum operator $P$. Together with the boost operator $K$, the operators $H$, $P$ (combined into $P^\mu$) and $K$ generate the 2-dimensional Poincar\'e symmetry group $IO(1,1)$, $[H,P] = 0$, $[K,H] = iP$ and $[K,P] = iH$ or $[P^+,P^-] = 0$ and $[K,P^\pm] = \pm iP^\pm$, with Casimir operator $P^2 = P^\mu P^\mu = H^2 - P^2 = P^+P^-$. The $IO(1,1)$ symmetry can be combined with an $SO(N)$ symmetry to obtain the $IO(1,N)$ space-time symmetry with as generators $H$, $P^i$, $K^i$ and $J^{[ij]}$ (combined into $P^\mu$ and $J^{\mu\nu}$, of course after also including discrete space- and time-reversal symmetries.

For massless excitations in 1D, right-movers (depending on $x^+$) are independent from left-movers (depending on $x^-$). Right- and left-handed fields satisfy $[P^-,\phi_R] = i\partial_+\phi_R = 0$ and $[P^+,\phi_L] = i\partial_-\phi_L = 0$. For massive fields left and right modes become coupled, while the other derivatives $[P^+,\phi_R] = i\partial_-\phi_R$ and $[P^-,\phi_L] = i\partial_+\phi_L$ acquire roles as (front form) canonical momenta~\cite{Dirac:1949cp}. For $M=0$ the fermion fields in $d=2$ satisfy $\gamma^-\xi_R = \gamma^+\xi_L = 0$ and $\xi_{R/L}$ are independent good fields~\cite{Kogut:1969xa}. Massive fermion fields satisfy the constraints $[P^-,\xi_R] = i\partial_+\xi_R = -iM\xi_L$ and $[P^+,\xi_L] = i\partial_-\xi_L = iM\xi_R$.

The $d=2$ Poincar\'e algebra in $E(1,1)$ can be extended to a supersymmetric algebra (for a review see Ref.~\cite{Martin:1997ns}) with anti-commuting fermionic operators $Q_{R/L}$,
\bea
&&
\{Q_{R},Q^\dagger_{R}\} = 2P^+, \quad
\{Q_{L},Q^\dagger_{L}\} = 2P^-,
\\&&
[P^\pm, Q_{R/L}] = 0, \quad [K, Q_{R/L}] = \pm \tfrac{1}{2}i\,Q_{R/L}.
\eea
Supersymmetry connects the fields, $[Q_{R/L},\phi_{R/L}] = \xi_{R/L}$ and $\{Q_{R/L},\xi^\dagger_{R/L}\} = [P^{\pm},\phi_{R/L}]$. We will first consider one type of $R/L$ fields ($N = 1$) in a single space dimension and then extend this to a set of (three) real scalar and real fermionic (Majorana) fields $\phi$ and $\xi$. If the masses are zero, right-movers ($R$) and left-movers ($L$) are independent degrees of freedom; for bosons a simple doubling; for fermions coinciding with right- and left-handed fermions. Starting for $N = 1$ with the Wess-Zumino model~\cite{Wess:1973kz} in two dimension,
\bea 
\lag & = & \tfrac{1}{2}\partial_-\phi_R\,\partial_+\phi_R 
+ \tfrac{1}{2}\partial_+\phi_L\,\partial_-\phi_L
\nonumber \\ && \mbox{} + \tfrac{i}{2}\,\xi_R\partial_+\xi_R 
+ \tfrac{i}{2}\xi_L\partial_- \xi_L - V(\phi,\xi)
\\ & = &
\tfrac{1}{2}\partial^\mu\phi_S\,\partial_\mu\phi_S 
+ \tfrac{1}{2}\partial^\mu\phi_P\,\partial_\mu\phi_P
+i\,\overline\psi\slash\partial\psi - V(\phi,\xi),
\nonumber\eea
(real) right and left fields for bosons can be combined into (real) scalar (CP-even) and pseudoscalar (CP-odd) fields $\phi_{S/P} = (\phi_R \pm \phi_L)/\sqrt{2}$. Real fermion fields can be combined in a (self-conjugate) spinor $\psi = (\xi_R,-i\xi_L)/\sqrt{2}$. 
Supersymmetry strongly restricts the interaction terms. The most compact expression is in terms of the scalar and pseudoscalar fields containing a mass term coupling left and right fields and a single Yukawa coupling that also governs the fermion-boson coupling,
\bea
V(\phi,\xi) & = & 
\tfrac{1}{2}(M+g\phi_S)^2(\phi_S^2 + \phi_P^2)
+\tfrac{1}{2}g^2\phi_P^2(\phi_S^2 + \phi_P^2)
\nonumber \\&& \mbox{} +\overline\psi(M + g\phi_S + g\phi_P\gamma^1)\psi 
+ \lambda\,F
\eea
using $\gamma_5 = \gamma^0\gamma^1$. The constraint is given by
\bea
\lambda\,F & = & 
\frac{\lambda}{4g}\left((M + 2g\phi_R\sqrt{2})(M + 2g\phi_L\sqrt{2}) 
- M^2\right)
\nonumber \\& = &
\frac{\lambda}{g}\left((g\phi_S+M/2)^2 - g^2\phi_P^2 - M^2/4\right).
\eea
Defining $M/2g \equiv v$, we introduce fields $\phi_S + v \equiv v\Phi_S$ and $v\Phi_P \equiv \phi_P$ which can be re-defined as $\Phi_S = \cosh \eta$ and $\Phi_P = \sinh \eta$ or if one likes one can use an {\em imaginary} representation for $\Phi_P$ by writing $\eta = i\theta$. The bosonic part of the potential including constraint becomes
\bea
V(\Phi) &=& 
\frac{v^2M^2}{2}\Phi_S^2\,\Phi_P^2 
= \frac{v^2M^2}{2}\Phi_S^2(\Phi_S^2-1) 
\eea
or $V(\Phi) = \tfrac{1}{8}\,M^2v^2\sinh^2(2\eta)$. 
Defining $\vert\Phi\vert^2 \equiv (\Phi_S^2 + \Phi_P^2)/2 = \Phi_R^2 + \Phi_L^2$ we have
$\vert\Phi\vert^2 = \cosh(2\eta)$ and we have
$\Phi_S^2 = \vert\Phi\vert^2 + 1/2$ and $\Phi_P^2 = \vert\Phi\vert^2 - 1/2$.
Looking at the minimum of the potential ($\eta = 0$ or $\theta = 0$) we see that the boson field acquires a vacuum expectation value which is right-left symmetric,
$\Phi_R = \Phi_L = \vert\Phi\vert = 1/\sqrt{2}$ (or $\Phi_S = 1$ and $\Phi_P = 0$). 
The real excitations around the vacuum are Majorana modes $\Psi = \Psi^c = (\xi, -i\xi)/\sqrt{2}$ and real scalar bosonic modes $\Phi_S/\sqrt{2} = \Phi = \Phi^c = (1+H)/\sqrt{2}$. Note that $\phi_S = v\,H$. 
The 1D pseudoscalar field $\phi_P$ can be identified as a vector field writing $i\partial_\mu\Phi_{R/L} = (i\partial_\mu \pm gA_\mu)\phi_S/\sqrt{2}$. In the ground state $A_\mu = 0$ and around the vacuum one has $A^\mu \approx \phi_P(n^\mu - \bar n^\mu)$ or $A^+ = -A^- \approx \phi_P$. This suggests working with a complex field $\Phi$ rather than left and right fields that are CP symmetric, $\Phi_R = \Phi_L^\ast$. For a single field a global $U(1)$ symmetry is not relevant and local symmetries don't lead to dynamics either, but taking multiple scalar fields the symmetry pattern becomes much richer.

\section{Extension to three fields}

The symmetric extension to $N$ real boson and fermion fields (we take $N = 3$), $\phi = (\phi_1,\phi_2,\phi_3)$, has interesting consequences for the dynamics, which is studied by looking at the possible fluctuations around the vacuum, in the symmetric basis $\langle \Phi\rangle^T = (1,1,1)/\sqrt{3}$. Including complex phases we consider $SU(3)$ fluctuations, although the lagrangian is only invariant under $SO(3)$ transformations of the fields, which is also the symmetry of the groundstate. We propose to use the $SO(3)$ symmetry in combination with inversion and time reversal symmetry, to extend the $d = 2$ Poincar\'e symmetry to a $d = 4$ Poincar\'e symmetry. Implemented in Weyl mode, the asymptotic fields become real representations of $IO(1,3)$ living in $E(1,3)$.  

\begin{figure}
\epsfig{file=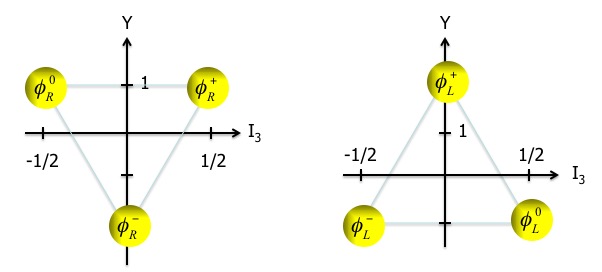,width=0.45\textwidth}
\caption{\label{basicbosons2b}
SU(3) quantum number assignments for bosonic excitations. The right-moving positively charged ($\phi_R^{+}$) and neutral ($\phi_R^0$) scalar fields are in an isospin doublet with $Y = +1$, while the right-moving negatively charged ($\phi_R^-$) field is in a weak isosinglet with $Y = -2$. The left-moving negatively charged ($\phi_L^-$) and neutral ($\phi_L^0$) fields are in an isospin doublet with $Y = -1$ while the left-moving positively charged ($\phi_L^{+}$) field is in a weak isosinglet with $Y = +2$. 
} 
\end{figure}

At this stage, part of the freedom in fluctuations around the vacuum has been incorporated. The already accounted for real $SO(3)$ rotations are identified with the subalgebra generated by the $SU(3)$ generators $\lambda_2$, $-\lambda_5$ and $\lambda_7$, constituting the algebra of the factor group of the subgroup $G^\prime = SU(2) \times U(1)$ with generators $\lambda_1/2$, $\lambda_2/2$, $\lambda_3/2$ and $\lambda_8/2$. This subgroup contains the $SU(3)$ Cartan subalgebra consisting of $I_3 = \lambda_3/2$ and $Y = \lambda_8\sqrt{3}$ that will serve as electroweak charge labels for weak isospin and hypercharge. Labelling the (massless) bosonic states using this Cartan subalgebra, gives fields $\phi_{R/L}^{[I_3,Y]}(\bm x,t)$ living in $E(1,3)$. For two fields this would have been just a $U(1)$ charge assigment. The basic bosonic starting point for the three fields and their electroweak quantum numbers is illustrated in Fig.~\ref{basicbosons2b}.

To account for the fluctuations around the vacuum, we look at the covariant derivatives,
\bea
E(1,1):&& 
i D_\mu\Phi^i = i\partial_\mu\Phi^i + g\sum_{a\in G} A_\mu^a (T_a)^i_j\Phi^j,
\label{covder-1}
\\
E(1,3):&&
iD_\mu\Phi^i = i\partial_\mu\Phi^i 
+ g\sum_{a\in G^\prime} A_\mu^a (T_a)^i_j\Phi^j.
\label{covder-3}
\eea
The first expression applies to fields in $E(1,1)$ and accounts for local $SU(3)$ gauge invariance. It involves eight (color) gauge fields also living in $E(1,1)$. The second expression is relevant for (asymptotic) fields in $E(1,3)$. Coupling for {\em real} continuous $SO(3)$ transformations the field and space rotations, there are no gauge fields for that part leaving only the complex transformations involving four (electroweak) gauge fields living in $E(1,3)$. 

The embedding of $SO(3)$ directions into $SU(3)$ is not unique. The discrete symmetry group $A_4$ governs the possible oriented embeddings. For singlet representations of this embedding group one can consider $SU(3)\supset SO(3) \times A_4 \times [SU(2)\otimes U(1)] \rightarrow  SO(3)\otimes [SU(2)\otimes U(1)]$, decoupling space-time and internal symmetries~\cite{Coleman:1967ad}. 
The unitary transformation matrix for these singlet states~\cite{Cabibbo:1977nk,Wolfenstein:1978uw,Ma:2001dn,Altarelli:2005yx} is the matrix $W$ that rotates the {\em symmetric} embedding of the vacuum into an {\em electroweak} embedding,
\be
W^\dagger\left\lgroup \begin{array}{c} \sqrt{1/3} \\ \sqrt{1/3} \\ \sqrt{1/3} \end{array}\right\rgroup = \left\lgroup \begin{array}{c} 0 \\ 1 \\ 0 \end{array}\right\rgroup  \ \mbox{for}\ W = \frac{1}{\sqrt{3}}\left\lgroup \begin{array}{ccc} 1 & 1 & 1 \\ \omega^2 & 1 & \omega \\ \omega & 1 & \omega^2 \end{array}\right\rgroup ,
\ee
where $\omega = \exp(i\,2\pi/3)$. Since the starting point only had $SO(3)$ as a symmetry group, the vacuum indeed is not invariant under $SU(2)\otimes U(1)$ transformations, but it is neutral for $Q = I_3 + Y/2$.
The symmetry pattern and its breaking thus is summarized as
\bea 
&&
IO(1,1) \otimes SU(3)
\nonumber\\&&\mbox{}\quad
\ \supset\ \underbrace{IO(1,1) \times SO(3)}_{IO(1,3)} \otimes 
\underbrace{SU(2)_I \otimes U(1)_Y}_{\rightarrow\ U(1)_Q}. 
\nonumber
\eea 
All bosons and fermions, however, still do originate as (finite dimensional) representations of the basic $SU(3)$ symmetry group, which will become important later. There are three families of particles corresponding to the singlets of $A_4$.
Going to three space dimensions the interaction changes from a confining potential to a $1/r$ (or Yukawa) potential between the (electroweak) charges, which thus can be {\em free}, in contrast to the (color) $SU(3)$ charges in one space dimension. 

\section{Electroweak sector}

After the introduction of the covariant derivatives, part of the potential is included in the term
\bea
D^\mu\Phi^\ast D_\mu\Phi & = & \partial^\mu\Phi^\ast \partial_\mu\Phi + \frac{N\,C_A}{2}\,g^2A^\mu A_\mu\,\Phi^\ast\Phi .
\label{massterm}
\eea
With $C_A(G=SU(3)) = 4/3$, the second term in Eq.~\ref{massterm} is precisely $-V(\Phi)$ and we are left with the $1+1$ dimensional QCD lagrangian (without a Higgs mass-term),
\be 
\lag = \tfrac{1}{2}\partial^\mu H\partial_\mu H -\frac{1}{4}\,F^{\mu\nu}F_{\mu\nu}
+ \overline\Psi (i\slash D - M - gv\,H)\Psi ,
\ee
but with a scalar field, which does not seem harmful~\cite{Kaplan:2013dca}. We will first consider the electroweak structure of the lepton sector before returning to that of the colored fermions (quarks).

For the second option of the covariant derivative (Eq.~\ref{covder-3}) we have $C_A(G^\prime) = 2/3$, the second term in Eq.~\ref{massterm} is only $-V(\Phi)/2$. This leaves the scalar field massive with $M_H = M/\sqrt{2}$. This is an experimentally interesting scenario for the standard model if the fermion mass is identified with the top quark mass $M = M_{\rm t}$.

For the bosons, we have (in principle arbitrarily) assigned right to the triplet and left to the anti-triplet. The fields can be rotated into a single scalar field with a nonzero vacuum expectation value as is done in the usual standard model treatment, even if they form triplets,
\bean
\Phi_L & = & \frac{1}{\sqrt{2}}\,\exp\bigl(-\tfrac{i}{2}\sum_{a=1,2,3}\theta^a \lambda_a\bigr)\left(
\begin{array}{c} 0 \\ 1 + H \\ 0\end{array}\right), 
\\
\Phi_R & = & \frac{1}{\sqrt{2}}\,\exp\bigl(+\tfrac{i}{2}\sum_{a=1,2,3}\theta^a \lambda_a\bigr)\left(
\begin{array}{c} 1 + H \\ 0 \\ 0\end{array}\right).
\eean 
The electroweak charges and corresponding generators of gauge transformations are identified with the $SU(2)_I \otimes U(1)_Y$ transformations but with a single coupling constant within $SU(3)$. The charged fields are neither $I_3 = \lambda_3/2$ or $Y = \lambda_8\sqrt{3}$ eigenstates but they are eigenstates of $Q = I_3 + Y/2$. The breaking of the $SU(2)_I\times U(1)_Y$ symmetry to $U(1)_Q$ after the choice of ground state being neutral, produces three massive and one massless gauge boson. As discussed in a slightly different context \cite{Weinberg:1971nd}, the $SU(3)$ embedding gives a weak mixing angle, $\sin\theta_W = 1/2$ after rewriting in 
\bea 
iD_\mu\Phi &=& i\partial_\mu\Phi + 
\frac{g}{2}\,\biggl(\sum_{i=1}^3 W_\mu^i\lambda_i + B_\mu\lambda_8\biggr)\Phi
\label{covsu3}
\eea
the neutral combination $gW_\mu^0 I_3 + (g/2\sqrt{3})B_\mu Y$ in terms of $Z^\mu$ and $A^\mu$.
One obtains (using the dimensionful coupling in $d=2$) $e = g/2$ and masses $M_W^2 = M^2/4$, $M_Z^2 = M_W^2/\cos^2\theta_W = M^2/3$ and $M_A^2 = 0$.
In zeroth order, the weak mixing is fine and the Higgs mass and gauge boson masses are related and they are of the right order with $M = M_{\rm t}$. Taking $v = M/2g = 1$, one even is tempted to compare $e/M = 1/4$ with $\sqrt{4\pi\alpha} \approx 0.3$. Besides providing a global zeroth order picture for electroweak bosons, we note that the left-right symmetric starting point also ensures custodial symmetry~\cite{Veltman:1977kh,Sikivie:1980hm}.

\begin{figure}[t]
\epsfig{file=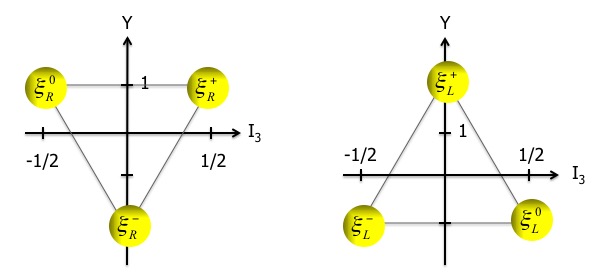,width=0.45\textwidth}
\caption{\label{basicfermions}
SU(3) quantum number assignments for fermionic excitations. The left-moving negatively charged ($\xi_L^{-}$) Dirac field and the neutral ($\xi_L^0$) Majorana field belong to an isospin doublet with $Y = -1$, while the left-moving positively charged ($\xi_L^+$) field is a weak isosinglet with $Y = +2$. The right-moving positively charged ($\xi_R^+$) Dirac field and neutral ($\xi_R^0$) Majorana field belong to an isospin doublet with $Y = +1$ while the right-moving negatively charged ($\xi_R^{-}$) field is a weak isosinglet with $Y = -2$. The asymptotic fields become right- and left-handed fields, $\xi_{R/L}^\pm \rightarrow e_{R/L}^\pm$ and $\xi_{R/L}^0 \rightarrow \nu_{R/L}^0$.
} 
\end{figure}

For the fermionic excitations, the starting $SU(3)$ triplets $\xi_R$ and anti-triplets $\xi_L$ in 1D match those of the bosons, implying the underlying supersymmetry of the elementary fermionic and bosonic $d=2$ excitations. Also in this case one fixes one direction for the $SU(3)$ representations (the $SO(3)$ embedding) and uses the (remaining) symmetry to fix the electroweak structure as an $SU(3)$ triplet or anti-triplet. The fermions then have electroweak charges corresponding to isospin doublets and singlets as shown in Fig.~\ref{basicfermions}. We already mentioned the possible role of the $A_4$ embedding symmetry in the family structure of fermions, which allows three independent families. Besides the matrix $W$ that transforms between symmetric and electroweak basis,
\bea
Q_{\rm s} &=& \frac{1}{\sqrt{3}}\left\lgroup \begin{array}{ccc} 0 & i & -i \\ -i & 0 & i \\ i & -i & 0 \end{array}\right\rgroup = W \left\lgroup \begin{array}{ccc} 1 & 0 & 0 \\ 0 & 0 & 0 \\ 0 & 0 & -1 \end{array}\right\rgroup W^\dagger.
\nonumber\\&&
\eea
one needs to transform Majorana fermions ($\xi_1$, $\xi_2$, $\xi_3$) into charged fermions ($\xi^+$, $\xi^0$, $\xi^-$), which we do by mixing $\xi_1$ and $\xi_3$ in the symmetric basis, such that
\bea 
Q_{\rm ew} = \left\lgroup \begin{array}{ccc} 0 & 0 & -i \\ 0 & 0 & 0 \\ i & 0 & 0 \end{array} \right\rgroup = V_Q \left\lgroup \begin{array}{ccc} 1 & 0 & 0 \\ 0 & 0 & 0 \\ 0 & 0 & -1 \end{array}\right\rgroup V_Q^\dagger .
\eea 
This shows that $Q_{\rm s} = U_{\rm HPS}\,Q_{\rm ew}\,U_{\rm HPS}^\dagger$ in which the tribimaximal mixing matrix~\cite{Harrison:2002er} appears, $U_{\rm HPS} = W V_Q^\dagger$
\bea
U_{\rm HPS} &=& \left\lgroup \begin{array}{ccc}
\sqrt{2/3} & \sqrt{1/3} & 0 \\ -\sqrt{1/6} & \sqrt{1/3} & - \sqrt{1/2} \\ -\sqrt{1/6} & \sqrt{1/3} & \sqrt{1/2} \end{array} \right\rgroup.
\eea
This looks like a promising zeroth order description for leptons providing arguments for the role of the discrete symmetry group $A_4$, which is a subgroups of both $SO(3)$ {\em and} $SU(3)$, in the structuring of families and the mixing matrices. The details of this, however, need to be worked out.

Finally note that (without looking at the role of the masses) the extension of 1D fermion fields leads to 3D 'good' light-front fields $\Psi = (\xi_R,-\epsilon\xi_L)$ with two-component spinors $\xi_{R/L}$ and $\epsilon = i\sigma^2$. The rotations are represented by $\bm J = \bm\sigma/2$, boosts by $\bm K = \pm i\bm\sigma/2$ for right and left fields, respectively (thus $n^\mu \rightarrow \sigma^\mu$ and $\bar n^\mu \rightarrow \bar\sigma^\mu$). The coupling of fermions to the pseudoscalar fields, combined into a 3D vector field, becomes the $\overline\Psi \slash A\Psi$ coupling. 


\section{Strong sector}

The fermionic modes $\xi$ can also just live in $E(1,1)$ and be arranged in three families of $SU(3)_C$ color triplets, which are identified as colored quarks but living in E(1,1) where color is confined via the instantaneous confining linear potential of the gauged $SU(3)$ symmetry. In order to study the electroweak structure of quarks (their {\em valence} nature) one has to study their interactions with the electroweak gauge bosons. We propose to do this by mapping the structure of the excitations into three spatial directions in a {\em frozen color} scheme in which we just consider fermions of one particular color (say $r$). Take the case of all $\xi_R$ states with color $r$ and all $\xi_L$ being $\bar r$. Taking a step back and looking at what was done in order to find leptons where the frozen colors were in essence space dimensions. The one-dimensional state would be labeled by a single momentum component, which is extended to states labeled by a 3-dimensional momentum vector in $E(1,3)$. For two space dimensions, the fermions could be labeled by their helicity in $E(1,2)$, charge eigenstates being ($\xi^-\xi^-$), ($\xi^+\xi^+$) and ($\xi^0\xi^0$). For leptons in three space dimensions $\xi_L^0$ was combined with ($\xi_L^0\xi_L^0$) to find an asymptotic charge eigenstate with $(Q,I_3) = (0,+1/2)$, which we already discussed as the left-handed Majorana neutrino $\nu_L$. For colored eigenstates we specify how states are 'viewed' in 3 dimensions by combining the (frozen) anti-red $\xi_L^0$ state with the (frozen) $rr$ combinations ($\xi_R^0,\xi_R^0$), ($\xi_R^+\xi_R^+$) or ($\xi_R^-\xi_R^-$). Then only the combination ($\xi_R^+\xi_R^+$) leads to acceptable $SU(3)$ quantum numbers (roots), being an asymptotic acceptable $SU_I(2)$ weak eigenstate with $I_3 = 1/2$, which has $U_Q(1)$ charge $Q = +2/3$, identified as the weak iso-doublet quark state $u_L$ with color $r$ belonging to a color triplet. Combining the (frozen) color $\bar r$ state $\xi_L^0$ with the (frozen) $r\bar r$ combination ($\xi_L^0\xi_R^0$), ($\xi_L^+\xi_R^+$) or ($\xi_L^-\xi_R^-$) gives only for ($\xi_L^-\xi_R^-$) an acceptable (frozen) color $\bar r$ state with $(Q,I_3) = (-2/3,0)$, the weak iso-singlet antiquark state $\bar u_L$. The full set of electroweak assignments of quarks as viewed in three space dimensions is shown in Table~\ref{content-fermions}. The resulting allowed $SU(3)$ quantum numbers are for each family a left-handed quark doublet and right-handed antiquark doublet and two singlets of opposite handedness. The way in which the electroweak structure emerges resembles the rishon model~\cite{Harari:1981uh}, but rather than having two fractionally charged preons ($V$ and $T$) in $d = 4$, our basic modes are charged or neutral preons living in $d = 2$. 
The family mixing would also for quarks originate from symmetries in fixing a direction, but in zeroth order there is only a single heavy quark, the top quark (with $M_t = M$), so the mixing would be trivial. But it is fair to say, that a complete mechanism for masses and mixing for quarks and leptons requires further study. 

\begin{table}
\begin{tabular}{|c|ccc|cc|c|c|c|}
\hline
{} {} & \multicolumn{3}{|c|}{\quad space\ \quad {} } & \multicolumn{3}{|c|}{electroweak} & {}charge {} & {}color{} \\
& {}\quad $L_{\ {}}$ \quad {} & {}\ ($T_1$  \ {} & {}\ $T_2$)  \ {} & {}\quad $I$\quad{} & {}\quad$I_3$\quad{} & $Y$ & $Q$ & $\underline{c}$ 
\\
\hline
$\nu_L$ & $\xi_L^0$ & $\xi_L^0$ & $\xi_L^0$ & 1/2 & +1/2 & $-1$ & 0 & $\underline{1}$
\\
$e_L^-$ & $\xi_L^-$ & $\xi_L^-$ & $\xi_L^-$ & 1/2 & $-1/2$ & $-1$ & $-1$ & $\underline{1}$
\\[0.1cm] \hline
$e_L^+$ & $\xi_L^+$ & $\xi_L^+$ & $\xi_L^+$ & 0 & 0 & +2 & +1 & $\underline{1}$ \\[0.1cm] \hline\hline
$\nu_R$ & $\xi_R^0$ & $\xi_R^0$ & $\xi_R^0$ & 1/2 & $-1/2$ & +1 & 0 & $\underline{1}$
\\
$e_R^+$ & $\xi_R^+$ & $\xi_R^+$ & $\xi_R^+$ & 1/2 & +1/2 & +1 & +1 & $\underline{1}$
\\[0.1cm] \hline
$e_R^-$ & $\xi_R^-$ & $\xi_R^-$ & $\xi_R^-$ & 0 & 0 & $-2$ & $-1$ & $\underline{1}$ \\[0.1cm] \hline\hline
$u_L$ & $\xi_L^0$ & $(\xi_R^+$ & $\xi_R^+)$ & 1/2 & +1/2 & +1/3 & +2/3 & $\underline{3}$
\\
$d_L$ & $\xi_L^-$ & $(\xi_R^0$ & $\xi_R^0)$ & 1/2 & $-1/2$ & +1/3 & $-1/3$ & $\underline{3}$
\\[0.1cm] \hline
$\overline{u}_L$ & $\xi_L^0$ & $(\xi_L^-$ & $\xi_R^-)$ & 0 & 0 & $-4/3$ & $-2/3$ & $\underline{3}^\ast$ 
\\[0.1cm] \hline
$\overline{d}_L$ & $\xi_L^+$ & $(\xi_L^0$ & $\xi_R^0)$ & 0 & 0 & +2/3 & +1/3 & $\underline{3}^\ast$ 
\\[0.1cm] \hline\hline
$\overline{u}_R$ & $\xi_R^0$ & $(\xi_L^-$ & $\xi_L^-)$ & 1/2 & $-1/2$ & $-1/3$ & $-2/3$ & $\underline{3}^\ast$
\\
$\overline{d}_R$ & $\xi_R^+$ & $(\xi_L^0$ & $\xi_L^0)$ & 1/2 & +1/2 & $-1/3$ & +1/3 & $\underline{3}^\ast$
\\[0.1cm] \hline
$u_R$ & $\xi_R^0$ & $(\xi_L^+$ & $\xi_R^+)$ & 0 & 0 & +4/3 & +2/3 & $\underline{3}$ 
\\[0.1cm] \hline
$d_R$ & $\xi_R^-$ & $(\xi_L^0$ & $\xi_R^0)$ & 0 & 0 & $-2/3$ & $-1/3$ & $\underline{3}$ 
\\[0.1cm] \hline\end{tabular}
\caption{\label{content-fermions}
Fermionic excitations with their assigments in $SO(1,1) \times SU(3)_C$ and 
$SO(1,3) \times SU(2)_I \times U(1)_Y$ symmetry schemes. The column labeled space $L$ indicates one of the  basic (left/right) modes of the $d = 2$ theory (the colored fermionic modes). The columns labeled $(T_1\ T_2)$ contain charge eigenstates of two basic modes that can be combined with the $L$-mode, giving allowed $I_3$ states within $SU(3)$. 
}
\end{table}

\section{Conclusions}

Concluding, instead of extending the standard model of particle physics, I have described an attempt to start at a more basic level with just a single space dimension ($d = 2$) and as starting point a fully supersymmetric set of three real preon fields describing bosonic and fermionic excitations. With this supersymmetric, superrenormalizable starting point, there is no naturalness or hierarchy problem. The $SO(3)$ symmetry of the classical ground state, including parity and time reversal, is then in Weyl mode realized as excitations living in 3D. The bosonic degrees of freedom are rearranged into the Higgs particle and the electroweak gauge bosons, while fermions are arranged in three families with two charged (Dirac) and one neutral (Majorana) lepton arranged in left-handed weak isospin doublets and singlets and corresponding right-handed antileptons. All these excitations appear as asymptotic states in 3D.
The excitations of the fields also can live in 1D. The $SU(3)$ gauge theory has an instantaneous confining interaction and no physical gauge degrees of freedom. But this is not how these degrees of freedom show up asymptotically. We argue that the quarks reveal themselves in 3D as good (front form) components of fractionally charged Dirac fields arranged in a lefthanded weak isospin doublet and two righthanded singlets (and corresponding right- and left-handed antiparticles). 

In this way a minimal scenario is created to obtain the standard model of particle physics with also in 3D {\em elementary} fields, while confinement of color is implicit. Most prominent is that it links the number of colors, families and space directions. The Higgs or top quark mass are the natural basic scales for wave-lengths of the one-dimensional excitations producing the right orders of magnitude for masses of top quark, Higgs particle and gauge bosons. There are many details that need to be investigated to see if the proposed scheme can be made consistent, the embedding mechanism for the family structure, the origin of mixing matrices, the emergence of the scale of QCD, etc.
The conjectures as put forward here will likely not invalidate the existing highly successful field theoretical framework for the standard model. Hopefully a more explicit treatment could provide ways to calculate its parameters. The 1D starting point for the strong sector also may provide insights why and to what extent descriptions like the AdS/QCD correspondence (see e.g.\ Ref.~\cite{deTeramond:2008ht}), collinear effective theories (see e.g.\ review in Ref.~\cite{Becher:2014oda}) or the many effective theories for QCD at low energies work. The link with the family structure might provide handles on universality breaking effects such as the 'proton radius puzzle'. The reason is that atomic Hydrogen involves all degrees of freedom of just {\em one} family while muonic Hydrogen is different in this respect. It could also be interesting to look at more (or maybe less) than three fields, which could be relevant in the context of the evolution of our universe into the world which above hadronic scales, i.e.\  the visible part at nuclear, atomic, molecular scales up to astronomical scales, is governed by three space dimensions.

\section*{Acknowledgements}

I acknowledge useful discussions with colleagues at Nikhef, in particular with Tomas Kasemets. 
This research is part of the FP7 EU "Ideas" programme QWORK (Contract 320389).

\bibliographystyle{apsrev}

 

\end{document}